\documentclass[useAMS,usenatbib,usegraphicx]{mn2e}
\usepackage{epsfig}
\usepackage{amsmath} 
\usepackage{rotating}           
\usepackage{color}     
\usepackage{graphicx}
\usepackage{times}
\usepackage{upgreek} 

\def\kms{km ${\rm s}^{-1}$}

\def\ch2{$\chi^2$}
\def\dg{$^{\circ}$}

\def\kms {\hbox{${\rm km\ s}^{-1}$}}

\def\ccm {$\hbox{{\rm cm}}^{-3}$}    
\def\scm  {$\hbox{{\rm cm}}^{-2}$}    

\def\MOLH {\hbox{${\rm H}_2$}}  

\def \AL {$\alpha $}     
\def \HI {H{\sc \,i}}
\def \WpHz {W Hz$^{-1}$}
\def\lapp{\ifmmode\stackrel{<}{_{\sim}}\else$\stackrel{<}{_{\sim}}$\fi}
\def\gapp{\ifmmode\stackrel{>}{_{\sim}}\else$\stackrel{>}{_{\sim}}$\fi}

\title[\HI\ 21-cm absorption and  impact parameter in galaxies]{A correlation between the \HI\ 21-cm absorption strength and impact parameter in external galaxies}

\author[S. J. Curran et al.]{S. J. Curran$^{1}$\thanks{Stephen.Curran@vuw.ac.nz},  S. N. Reeves$^{2,3,4}$, J. R. Allison$^{2}$ and E. M. Sadler$^{3,4}$\\
$^{1}$School of Chemical and Physical Sciences, Victoria University of Wellington, PO Box 600, Wellington 6140, New Zealand\\
$^{2}$CSIRO Astronomy and Space Science, PO Box 76, Epping NSW 1710, Australia\\
$^{3}$Sydney Institute for Astronomy, School of Physics, University of Sydney, NSW 2006, Australia\\
$^{4}$ARC Centre of Excellence for All-sky Astrophysics (CAASTRO)}

\begin{document}

 \date{Accepted ---. Received ---; in original form ---}

\pagerange{\pageref{firstpage}--\pageref{lastpage}} \pubyear{2016}

\maketitle
\label{firstpage}
\begin{abstract}
  By combining the data from surveys for \HI\ 21-cm absorption at various impact parameters in near-by galaxies, we
  report an anti-correlation between the 21-cm absorption strength (velocity integrated optical
  depth) and the impact parameter.  Also, by combining the 21-cm absorption strength with that of the emission, giving the
  neutral hydrogen column density, $N_{\text{\HI}}$, we find no evidence that the spin temperature of the gas
  (degenerate with the covering factor) varies significantly across the disk. This is consistent with the uniformity of
  spin temperature measured across the Galactic disk.  Furthermore, comparison with the Galactic $N_{\text{\HI}}$
  distribution suggests that intervening 21-cm absorption preferentially arises in disks of high inclinations (near
  face-on).  We also investigate the hypothesis that 21-cm absorption is favourably detected towards compact radio
  sources. Although there is insufficient data to determine whether there is a higher detection rate towards quasar,
  rather than radio galaxy, sight-lines, the 21-cm detections intervene objects with a mean turnover frequency of
  $\left<\nu_{_{\rm TO}}\right>\approx5\times10^{8}$ Hz, compared to $\left<\nu_{_{\rm TO}}\right>\approx1\times10^{8}$
  Hz for the non-detections. Since the turnover frequency is anti-correlated with radio source size, this does indicate
  a preferential bias for detection towards compact background radio sources.
\end{abstract}

\section{Introduction} 
\label{intro}
 
\begin{figure}
\centering \includegraphics[angle=-90,scale=0.55]{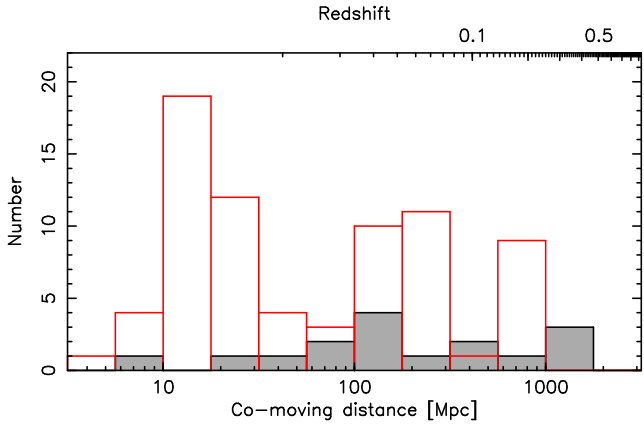}
\caption{The redshift distribution of the galaxies  searched for \HI\ 21-cm at various impact parameters.
The filled histogram shows the 21-cm detections and the unfilled the non-detections. The galaxies of \citet{rsa+15,rsa+16} are selected from the \HI\ Parkes All-Sky Survey (HIPASS, \citealt{ksk+04}) 
and so have redshifts of $z < 0.04$.}
\label{distbn}
\end{figure}

Study of the cool neutral gas over various redshifts gives insight into the evolution of the star forming reservoir and
is a science goal of the forthcoming Square Kilometre Array (SKA, \citealt{msc+15}). The distribution of cool neutral
hydrogen (\HI) with galactocentric radius in external galaxies is crucial in interpreting the data from the forthcoming
surveys of \HI\ 21-cm absorption with the SKA and its pathfinders.\footnote{For example, the {\em First Large Absorption
    Survey in \HI} (FLASH) on the Australian Square Kilometre Array Pathfinder (ASKAP), of which the {\em Boolardy
    Engineering Test Array} has already detected redshifted \HI\ 21-cm (at $z=0.44$, \citealt{asm+15}).} To this end, we
\citep{rsa+15,rsa+16} have undertaken a survey of 21-cm absorption of background radio sources at various impact
parameters in the disks of intervening, nearby galaxies ($z < 0.04$, Fig.~\ref{distbn}).

For impact parameters of $\rho\lapp20$ kpc, the \HI\ detection rate is approximately 50 per-cent
\citep{gsb+10,bty+11,bmh+14,sgr+13}. However, \citet{rsa+15} find a significantly lower detection rate (4 per-cent),
which they attribute to the unbiased nature of their sample.  That is, no target pre-selection based upon impact
parameter nor the nature of the background source.  For a given sight-line, a background radio galaxy could result in a
lower covering factor, reducing the observed optical depth in comparison to a quasar (see Sect.
\ref{ebss}). 
 
Although the gas density in a galaxy is known to decrease with galactocentric radius (e.g. \citealt{too63}),
none of the previous surveys have found convincing evidence of a decrease in 21-cm absorption strength with impact
parameter (\citealt{gsb+10,bty+11,gsn+13,zlp+15}). This is despite evidence of an anti-correlation between the Lyman-\AL\ and C{\sc iv}
equivalent widths and the impact parameter being well established in the optical/UV bands
(e.g. \citealt{lbtw95,tls98,clwb01,rbt+11,bht+15}). By adding the latest results \citep{rsa+15,rsa+16,dgso16} to the
previously published data and incorporating the limits, obtained from the non-detections, here we
report the first clear evidence of an anti-correlation between the 21-cm absorption strength and the impact parameter.

\section{Data, results and analysis}
\subsection{\HI\ 21-cm absorption strength and impact parameter}
\subsubsection{Raw impact parameters}

There are now 90 spectral line observations of radio source sight-lines through the disks of galaxies (summarised in Table~\ref{table1}).
\begin{table*}  
\centering 
\begin{minipage}{180mm}
  \caption{Summary of the searches for \HI\ 21-cm absorption at various impact parameters in external galaxies.  In the
    interest of consistency we quote the NED name of the galaxy, followed by its redshift. ``Type'' designates
    whether the search targetted quasar--galaxy pairs (P) or metal-line systems (M). For ease of matching, the sight-line is
    named as in the reference, followed by its NED classification -- galaxy (G), QSO (Q) or unclassified
    radio source (U). This is followed by the velocity  integrated optical depth (or the $3\sigma$ limit) of the 21-cm absorption,
    the impact parameter, the search reference and finally the blue and far-infrared luminosities as obtained from the  
    photometry (see main text). \label{table1}}
\begin{tabular}{@{}l c  cc  c c r  c r c  c@{}}  
\hline
Galaxy (NED)                                     & $z_{\rm gal}$  & Type & Sight-line            &  Class&  $N_{\text \HI}$  & $\int\!\tau dv$  & $\rho$ & Ref. & $L_{\rm B}$ & $L_{\rm K}$\\
                                                          &                    &        &                              &         &  [$10^{20}$\, \scm]  &[\kms]                & [kpc]     &      & \multicolumn{2}{c}{[$\log_{10}$\WpHz]} \\
\hline
CGCG\,045--091                             & 0.032465  &  P &  134528.765+034720.09& Q         & ---              &  $<0.08$ & 53.3 & B11  &  21.54  & 22.42\\
CGCG\,244--046                             & 0.032092 &  P& 125247.588+474042.81& U            &   ---            &  $<0.06$ & 18.1  & B11 & 21.55  & 22.61\\
  ...                                                    &  ...               & ... & 125249.326+474042.19 & Q      &    ---           &  $<0.10$ & 15.2   &  B11  & ... & ...\\
ESO\,150--G005                                &  0.004790   & P & C-ESO150-G005-1  &   U &   ---                      & $<14.1$     &  18.7 & R15  & 20.60 &  ---\\
...                                                      & ...                & P & C-ESO150-G005-2   & U   &                     ---  & $<23.3$     &   25.1 & R15 & ... & ...\\
ESO\,300--G014                               &0.003186    &  P & C-ESO300-G014-1  &U  &       $1.4\pm0.1$  &$<1.22$        &  11.6 & R16 & 20.62&  21.78\\
...                                                       &    ...            & .. & C-ESO300-G014-2   &U&       $8.1\pm0.2$ & $<1.63$         &    6.1 & R16 & ... & ...\\

ESO\,345--G046                                &   0.007168  & P & C-ESO345-G046      &   U            &    $28\pm12$ &  $<2.90$     & 13.6 & R15 & 21.16& ---\\    
ESO\,357--G012                                 & 0.005227   & P& C-ESO357-G012-1a  & U&   $1.0\pm0.2$  &$<3.49$        &   20.0 &R16 &  20.82 &  21.69\\
...                                                       &    ...            & ... & C-ESO357-G012-1b & U &  $1.0\pm0.2$     &$<11.7$        &   18.8 & R16 & ... & ...\\
ESO\,363-G015                                &0.004216     & P & C-ESO363-G015-1   & U &       ---     &$<1.90$        &   18.1 & R16 &   20.77 & 19.28\\
...                                                     &    ...  &           ... & C-ESO363-G015-2a  & U &       ---              &$<6.09$        &   17.0& R16 & ... & ...\\
...                                                      &    ...  &        ... & C-ESO363-G015-2b  & U &   ---    &$<6.81$       &   17.6& R16 & ... & ...\\  
ESO\,400-G012                                & 0.026999 &  M  & 2030--370                   & Q      &  $2.7\pm1.3$ &   0.22  & 10.0  & C92    & 21.48  & 21.79\\
ESO\,402-G025                                &0.008577   & P& C-ESO402-G025-1  &  U &     $5.2\pm1.1$   & $<1.18$    & 17.7   &R15 & 20.46 & 20.57 \\
...                                                       &    ...           & P  &C-ESO402-G025-2   & U &      $9.3\pm1.0$          & $<2.88$    & 16.9  & R15 & ... & ...\\ 
ESO\,576-G069                                &0.017809 &   M  & 1327--206         & Q     &  $0.6\pm0.2$ &  0.20  & 13.9  & C92    & 21.48 & 22.24 \\
IC\,1914                                            & 0.003432    & P& C-IC1914-1a             & U                 & ---     &$<1.34$       &   17.2 & R16 & 20.63  & 19.76 \\
 ...                                                       &    ...  &       ... & C-IC1914-1b            & U &    ---                 &$<1.91$         &   16.9 & R16 & ... & ...\\ 
IC\,1954                                           &  0.003542     &  P & C-IC1954                  & U              & ---  & $<0.48$  &  10.7 & R15 &  20.50  & 21.40\\
IC\,4386                                          & 0.006278    &  P & C-IC4386-1               & U &      ---     &$<0.97$        &   21.5  & R16 & 21.57  & 20.29\\
  ...                                                   &  ...               &  ...  & C-IC4386-2               & U &   $3.2\pm0.2$   & $<3.91$       &  26.6 & R16 & ... & ...\\
IRAS\,02483+4302                           & 0.051440 & M  &0248+430                      &  Q     & ---               &  0.26   & 15.0 & H04   & ---  & 22.69 \\
\protect[KAC2002]\,A                              & 0.437       &  M  & 1243--072                   &  Q      &   ---                     & 0.75 & 11.6  &  K02    & ---& ---\\
2MASX\,J08495751+5108416         & 0.312  & M  & J0849+5108  & Q & --- &                     $0.95$   & 14  & G13 &   21.87 &  22.02\\
2MASX\,J13253523+4953246          & 0.047829  & P & 132534.240+495348.47 & U         &   ---            &  $<0.39$ & 23.9 & B11 &  21.12 & 22.20 \\
  ...                                                    & ...               & ... &  132534.565+495342.26 & Q     &  ---              &  $<0.64$ & 17.5 & B11& ... & ...\\
2MASX\,J13540065+5650007          & 0.095513   & M &  4C\,+57.23 & U &---            &$<1.17$    &  11 & Z15 & 21.61 &  22.38 \\
NGC\,2188 & 0.002492                   & P &C-NGC2188-1a          & G &   $1.5\pm0.2$   & $<6.25$       &  7.7 &R16 & 20.93  & 20.21\\
...                                                     &   ...  &...& C-NGC2188-1b   & G  &   $1.5\pm0.2$    &$<8.37$       &  8.0 & R16 & ... & ...\\
NGC\,0628                                      & 0.002192  &  P   &  0131+154 	              & Q       &  ---               &$<0.15$  &  94.1   & C90  &    21.18 & 21.54 \\
NGC\,0660                                      & 0.002835  & P    &  0139+132                    & U &   ---             & $<0.067$ & 51.9   &  C90    &20.17  & 21.96 \\
NGC\,1249                                     &0.003576     & P & C-NGC1249-1           & U &   $2.9\pm0.7$  &$<4.64$   &  13.2 & R16 & 21.18 &  20.67 \\
  ...                                                  &    ...             & ... & C-NGC1249-2            & U &    ---     &$<5.28$       &  19.6  & R16 & ... & ...\\
  ...                                                  &    ...             & ... &  C-NGC1249-3a         & U     & ---     &$<9.54$       &  25.0 & R16 & ... & ...\\
  ...                                                  &    ...             & ... & C-NGC1249-3b         & U     & ---     &$<10.5$       &  25.5 & R16 & ... & ...\\
NGC\,1566                                      & 0.005017    & P & C-NGC1566-1           & U&     $130\pm19$       &$<3.30$       &  17.2 & R16 & 21.83 & 22.01 \\
...                                                         &    ...  & ... &     C-NGC1566-2a         & U  &    $6.6\pm0.9$    &$<6.16$       &  25.5  & R16 & ... & ...\\
...                                                        &    ...  & ... & C-NGC1566-2b          & U  &    $7.7\pm0.9$  &$<13.42$       & 21.7 & R16 &... & ...\\
NGC\,3067$^*$                             & 0.004923 &  M   & 0955+326                    &  Q     &$0.8\pm0.4$   & 0.12  & 11.1 &  C92    & 21.18 & 21.48 \\
NGC\,4138                                      & 0.002962  &  P   &  3C\,268.4                    & Q       &   ---                 & $<0.08$   & 10.5   &  H75  & 20.88 &  21.57 \\
NGC\,4651                                      & 0.002628  &  P   &  3C\,275.1 	             & Q        &  --                &  $<0.010$ &  12.1 &  C90  &  21.21 & 21.60 \\
NGC\,5156                                      & 0.009967   &P &  C-NGC5156            &   U &    $14.3\pm2.2$    & $1.02$      &  18.5  & R16 & 22.15 & 22.43\\
NGC\,5832                                      & 0.001491  &  P   &  3C\,309.1                    & Q       &   $1.0$               & $<0.014$ & 11.2   & H75   & 20.23   & ---\\ 
NGC\,6503                                      & 0.000083  &  P   &  1749+70.1                  & Q       &  ---             & $<0.01$   & 10.3   &  B88 & 20.80  & 20.71 \\
NGC,7162A                                      & 0.007569    & P& C-NGC7162A-1       &   U &    $4.4\pm1.3$     & $<2.25$      &  16.1 & R16 & 21.22   & ---\\
....                                                     &    ...    &        ...& C-NGC7162A-2      &  G  &    ---                      & $<4.01$    &   46.7  & R16 &  ... & ...\\
NGC\,7412                                      & 0.005704   &  P & C-NGC7412-1           & U           &       $6.6\pm0.7$  &$<2.08$    &12.9 & R15 & 21.46 & 21.48\\
...                                                    & ...            &    ...  & C-NGC7412-2         & U            &   $6.6\pm0.7$     &  $<2.63$    & 12.6 & R15 & ... & ...\\
NGC\,7413$^*$                             & 0.032489  &  P   &  3C\,455	   	             & Q        & ---               & $<0.026$ & 14.9   &  C90 & 21.84 & 22.53\\
NGC\,7424                                      & 0.003132   &  P &  C--NGC7424             &  Q         &    $103\pm14$   & $<0.99$    & 10.0 & R15 & 21.02  & 20.62 \\
NGC\,7490                                      & 0.020724  &  P   &  2304+32B	             & U        & ---            & $<0.10$   & 204.8 &  C90 & ---& 22.53\\   
PGC\,016074                                  &  0.066046  & P    & 0446--208C                & Q        & ---               &    0.21       & 15.4   & C92  &  ---  & ---\\ 
...                                                    &  ...              & ...   & 0446--208E                & Q        & ---                &    0.32       & 37.1   & C92   &  ... &  ...  \\  
...                                                    &  ...              & ...   &  0446--208W              & Q        & ---               &    0.63       & 35.2   &  C92  &... &  ... \\ 
SDSS\,J074842.58+173450.6  & 0.052822          & P & 074841.773+173456.82 & Q &  ---   &  $<0.10$ & 13.5 & B11 & 21.18  & 22.15\\ 
           ...                                  & ...                      & ...  & 074841.786+173512.20 & U      &  ---        &  $<0.19$ &  25.1  & B11 & ... & ...\\
           ...                                  & ...                      &...   &074842.084+173443.37 & U  &  ---        &  $<0.26$ & 10.4 & B11 & ... & ...\\

\hline
\end{tabular}
\end{minipage}
\end{table*} 
 \begin{table*}  
\centering 
\begin{minipage}{180mm}
\addtocounter{table}{-1}
\caption{\em Continued}
\begin{tabular}{@{}l c  cc  c c r  c r c c @{}}  
\hline
Galaxy (NED)                                     & $z_{\rm gal}$  & Type & Sight-line            &  Class&  $N_{\text \HI}$  & $\int\!\tau dv$  & $\rho$ & Ref. & $L_{\rm B}$ & $L_{\rm K}$\\
                                                          &                     &        &                              &         &  [$10^{20}$\, \scm]  &[\kms]                & [kpc]     &      & \multicolumn{2}{c}{[$\log_{10}$\WpHz]} \\
\hline
SDSS\,J082153.75+503125.7         & 0.1835       & P    & J0821+5031                & Q    &    ---           & 	$<0.38$ & 15.9     & G10   &  --- & ---\\
...                                                    & ...               & P     & 082153.833+503120.57 & Q  &  ---        &  $<0.16$ & 16.0 & B11 & ... & ...\\
SDSS\,J084912.42+275740.4          & 0.1948    & P  & 084914.282+275729.90 & Q   &  ---       &  $<0.02$ & 86.7 & B11 &  21.82  & --- \\
SDSS\,J084957.48+510842.3         & 0.073485   & P   & J0849+5108                 & Q         &    ---       & 	$<0.08$ &  19.4 & G10   & 21.87  &  22.02\\
SDSS\,J085519.04+575140.7       & 0.026003 &   M& GB6\,J0855+5751 & G & ---            &$<0.75$    &  9.5 & Z15 & 20.49  & ---\\
SDSS\,J102257.92+123439.1       & 0.1253         & P&  102258.415+123426.26 & Q &  ---                 &  $<0.03$ & 33.1  & B11 & 21.46  &  22.37 \\
   ...                                                &...                 & P  & 102258.552+123439.94 & U   &  ---               &  $<0.13$ & 20.9  & B11 & .... & ...\\
SDSS\,J104257.74+074751.3           & 0.03321     & M   &J104257.74+074751.3 & Q        &  ---             &  $0.19$     & 1.7      &B10    &   --- & ---\\   
...                                                      & ...                &P    & 104257.598+074850.60 &Q    &  ---               &  $0.04962$ & 1.7 &  B11 & ... & ...\\
SDSS\,J010643.94-103419.3           & 0.146           & P & 010644.15-103410.5 & Q  &  ---       &  $<0.02$ & 23.4 & B11 & 21.69  &  22.54 \\
SDSS\,J110736.60+090114.7           & 1.22823 &   P& 110736.607+090114.72 & Q &   ---       &                  $<0.18$ & 8.1 & B11 & --- & ---\\
SDSS\,J111025.09+032138.8           & 0.030115 &   P  &  J1110+0321C             &  Q       &    ---            & $<2.8$ &  11.2 & G10 & 21.17 & ---\\
...                                                      &  ...                & ... & J1110+0321E              &  Q       &    ---             &  $<0.13$ &  22.5 & G10& .... & ...\\ 
...                                                      &  ...                & P  & J1110+0321W             &  Q       &    ---             &  $<0.08$ &  15.3 & G10 & .... & ...\\  
SDSS\,J124157.26+633237. 6          & 0.143         & P    &  J1241+6332C             & Q        &    ---            &  $2.90$ &  11.0 & G10 & --- & 22.49\\ 
...                                                      &  ...              & ...   & 1241+6332E               &  Q       &    ---            &  $<2.50$ &  34.0& G10 & ... & ...\\  
...                                                      &    ...            &...    & J1241+6332W             &  Q       &    ---            &  $<0.62$ &  53.0& G10 & ... & ...\\  
SDSS\,J122847.72+370606.9           & 0.138336  & P     & J1228+3706                &  Q        &    ---           &  $<0.07$ &  15.0& G10 & --- & ---\\ 
SDSS\,J132839.89+622136.0           & 0.0423      &  P  & 132840.599+622136.65  & Q     & ---              &  $<0.04$ & 4.2 & B11 &   20.46 & ---\\
SDSS\,J141629.25+372120.4           &  0.0341     &  P  & 141631.039+372203.01 & Q       &  $0.17^{\dagger}$  &  $<0.16$ & 32.4 & B11 & --- & ---\\
 ...                                                     & ...              & ... & 141630.672+372137.09 & U       &  ...             &  $<0.16$ & 16.2 & B14 & ... & ...\\
SDSS\,J144304.53+021419.3           & 0.371503  &  M  & J1443+0214 & G                         & ---             & $3.4$ & $<5$  &G13 &  21.75 & ---\\
SDSS\,J160659.13+271642.6           & 0.046199 & P   & 160658.315+271705.86 & Q      &   $0.32^{\dagger}$&             $<0.03$ & 23.3& B11 & 20.95 & ---\\
SDSS\,J163956.38+112802.1            & 0.08         & M &  J163956+112758 & G  &  ---                             & $15.7$ & 4  &S13 &  21.23  & 22.23 \\
UG\,00439                                          & 0.017669 & P & UM\,266 -- L1 & Q& $1.77\pm0.19^{\dagger}$      &$<0.07$    &  24.7 & D16& 21.75  & 22.06\\
...                                                         &    ...        & ... &  UM\,266 -- C&           ...&    ---     & ---    &  24.8 & D16&  ... & ...\\
...                                                         &    ...        & P & UM\,266 -- L2&...  &           ---     &$0.08$    &  25.7 & D16 & ... & ...\\
UGC\,07408                                      & 0.001541 &P    & 122105.480+454838.80 & U         &   $4.6^{\dagger}$        &  $0.11$ & 3.3 & B11   & 19.80 &  20.63\\
...                                                       & ....            &...&  122106.854+454852.16 & Q      &   $4.6^{\dagger}$         &  $1.53$ & 2.8 & B14 & ...& ...\\
...                                                       & ...             &... & 122107.811+454908.02 & Q     & $4.6^{\dagger}$            &  $<0.47$ & 2.6 & B14 & ...& ...\\\
UGC\,12081                                    & 0.038817  & P   &  2231+0953	              &  Q         & ---              & $<0.10$   & 92.3   & C90 & ---& 22.80 \\             
\hline
\end{tabular}
{References: H75 -- \citet{hb75}, B88 -- \citet{bdkb88}, C90 -- \citet{cs90}, C92 -- \citet{cv92}, K02 -- \citet{kac02}, H04 -- \citet{hc04}, B10 -- \citet{bty+10}, 
G10 -- \citet{gsb+10}, B11 -- \citet{bty+11}, G13 -- \citet{gsn+13}, 
S13 -- \citet{sgr+13}, B14 -- \citet{bmh+14}, R15 -- \citet{rsa+15}, Z15 -- \citet{zlp+15}, D16 -- \citet{dgso16}, R16 -- \citet{rsa+16}.\\
Notes: 
$^*$Also observed by \citet{hb75}. 
$^{\dagger}$Obtained from the quoted integrated flux density of the \HI\ emission or the quoted \HI\ mass.}
\end{minipage}
\end{table*} 
In order to fully utilise these data, we include the upper limits to the optical depths (from the non-detections), via
the {\em Astronomy SURVival Analysis} ({\sc asurv}) package \citep{ifn86}.  These are added to the 16 detected
sight-lines as censored data points.  For the bivariate data, a generalised non-parametric Kendall-tau test gives a
probability of $P(\tau) = 9.39\times10^{-4}$ of the observed $\int\tau dv$--$\rho$ anti-correlation correlation arising
by chance, which is significant at $S(\tau) = 3.31\sigma$, assuming Gaussian statistics (Fig. \ref{N-impact}, top
panel).
\begin{figure}
\centering \includegraphics[angle=-90,scale=0.5]{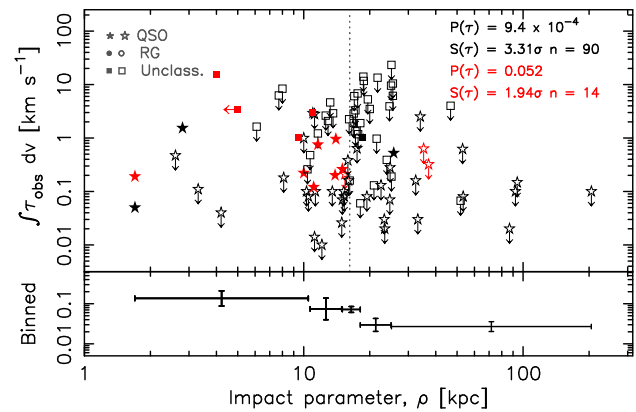}
\caption{The velocity integrated optical depth versus the impact parameter for all of the published searches. The shapes
  show the classification of the background source: star -- quasar/Quasi-Stellar Object (QSO), circle -- Radio Galaxy
  (RG) and square -- unclassified radio source, with the coloured symbols representing the known metal-line absorbers.
  The downward arrows signify the $3\sigma$ upper limits and the dotted vertical line shows where the sample is split in
  half with 45 sight-lines each at $\rho<16.2$ kpc and $\rho>16.2$ kpc.  The probability of the correlation arising by
  chance, $P(\tau)$, and the corresponding significance, $S(\tau)$, is shown in the top right for both the whole sample
  (black text) and the known metal-line absorbers only (coloured text). The bottom panel shows the binned values,
  including the limits via the Kaplan--Meier estimator, in equally sized bins. The horizontal error bars show the range
  of points in the bin and the vertical error bars the $1\sigma$ uncertainty in the mean value}
\label{N-impact}
\end{figure} 

As seen from Table \ref{table1}, only 14 of the 90 sight-lines were originally targetted as known metal-line
absorbers, although these drive the correlation giving a $S(\tau) =
1.94\sigma$ significance from just 14 data points. However, this is due to the metal-line absorbers
constituting 11 of the 16 detections, which define the parent population against which the limits are quantified,
thus requiring their inclusion.

The limits may also be included in the binned univariate data to give the mean  $\int\!\tau dv$ value of each bin, via  
the Kaplan--Meier estimator which gives a maximum-likelihood
estimate based upon the parent population \citep{fn85}.  Presenting the data thus, it is visually apparent that the absorption
strength does decrease with increasing impact parameter (Fig. \ref{N-impact}, bottom panel).

In addition to the generalised Kendall-tau test and the Kaplan--Meier estimates, we split the data in half via the
median impact parameter: At $\rho<16.2$ kpc, there are 14 detections and 31 non-detections, that is a 31.1 per-cent
detection rate. Applying this to the $\rho>16.2$ kpc bin, gives a binomial probability of $1.16\times10^{-5}$ of
obtaining two detections or fewer out of 45 sight-lines, which is significant at $4.37\sigma$.

\subsubsection{Normalised impact parameters}

Although by the inclusion of the limits, we see a decrease in absorption strength with impact parameter, the sample is
heterogeneous, comprising a variety of background source classifications (discussed in Sect. \ref{ebss}), in addition to
potential differences in the foreground absorbing galaxies. These may arise from the search strategy of each individual
survey, whether an unbiased survey towards galaxy--quasar pairs or the selection of {\em known} metal-line
absorbers.\footnote{Although this does not preclude any of the 76 designated as pairs as being potentially metal-rich --
  11 have been detected in Ca{\sc ii} absorption (see \citealt{cv92,hc04,gsb+10}), although a further nine have been
  searched and not detected (see \citealt{cs90,bty+10,gsb+10}).}  The detection rate of the former is just 7\%, compared
to 79 per-cent for the latter, with these also tending to be at low impact parameters (Fig. \ref{N-impact}).  Such
metal-line absorbers also comprise the foreground galaxies in the optical surveys which exhibit an equivalent
width--impact parameter anticorrelation (Sect. \ref{intro}).

Until all of the sample is observed and their metallicities quantified, this possible bias cannot be directly
removed. However, both metallicity and equivalent width appear to be correlated with galaxy size
(\citealt{ell06,ctp+07,mcw+07}) and so a potentially wide range of metallicities is expected to reflect a large range of
galaxy sizes.  This could have the effect of a given impact parameter, which lies well within the disk for an $L^*$
(high metallicity) galaxy, being located well outside the disk of a dwarf galaxy. In order to account for this, we
normalise each impact parameter by the luminosity of the galaxy: As described in \citet{cwsb12}, for each galaxy we
obtain the photometry from NASA/IPAC Extragalactic Database (NED), the Wide-Field Infrared Survey Explorer (WISE) and
the Two Micron All Sky Survey (2MASS), correcting each value for Galactic extinction \citep{wil13} before fitting to the
required wavelength in the galaxy rest-frame.

\begin{figure}
\centering \includegraphics[angle=-90,scale=0.5]{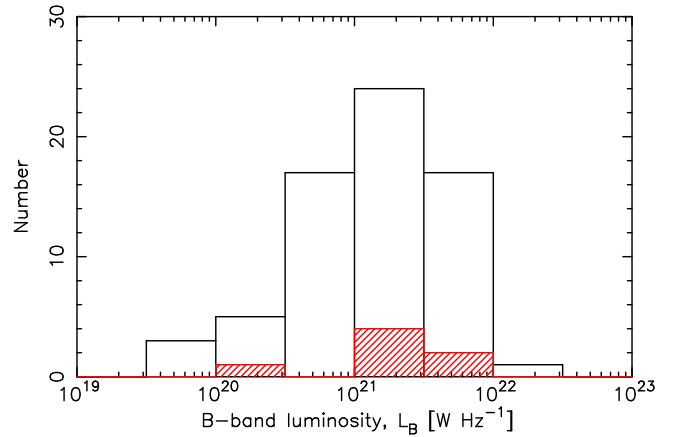}
\caption{The distribution of the blue-band luminosities of the absorbing galaxies, where sufficient photometry is
  available. The hatched histogram shows the metal-line searches and the unfilled the galaxy--quasar pair searches.}
\label{L-blue}
\end{figure}

We summarise the derived blue-band luminosities in Fig.~\ref{L-blue}.
Using these to normalise the impact parameter, via the same $(\rho/h)\times(L_{\rm B}/L_{\rm B}^*)^{-0.46}$ correction as \citet{clwb01},
\begin{figure}
\centering \includegraphics[angle=-90,scale=0.48]{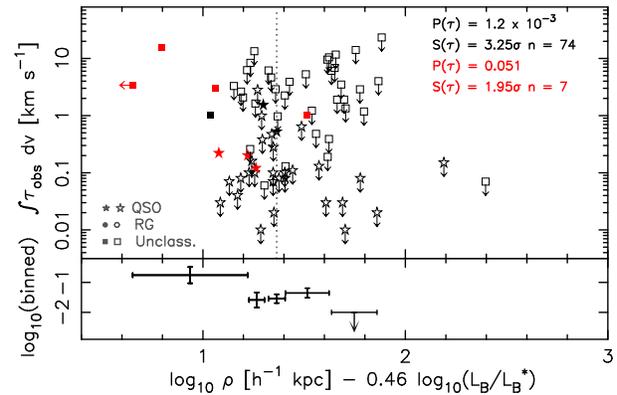}
\caption{As Fig. \ref{N-impact} but with the impact parameter normalised by the B-band luminosity (cf. \citealt{clwb01}).}
\label{N-blue}
\end{figure} 
we find the absorption strength--impact parameter anti-correlation  retains a similar significance, although with 16
fewer data points. As before, we can also compare detection rates above and below the median (normalised) impact
parameter: At $(\rho/h)\times(L_{\rm B}/L_{\rm B}^*)^{-0.46} <18.7$ kpc, there are 9 detections and 28 non-detections,
that is a 24.3 per-cent detection rate. Applying this to the $>18.7$ kpc bin, gives a binomial probability of
$4.28\times10^{-4}$ of obtaining two or fewer detections out of 37 sight-lines, which is significant at $3.52\sigma$.

\citet{clwb01} also normalise the impact parameter by the near-infrared luminosity, since this is less susceptible to extinction and
irregularities caused by star formation, providing a better tracer of the total stellar mass. 
\begin{figure}
\centering \includegraphics[angle=-90,scale=0.48]{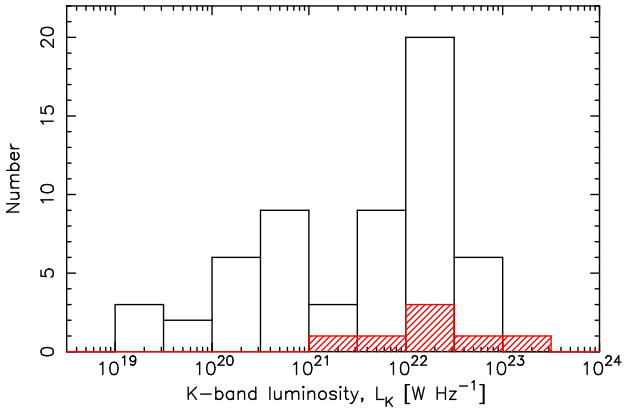} 
\caption{As Fig. \ref{L-blue}, but for the near-infrared luminosity.}
\label{L-K}
\end{figure} 
\begin{figure}
\centering \includegraphics[angle=-90,scale=0.5]{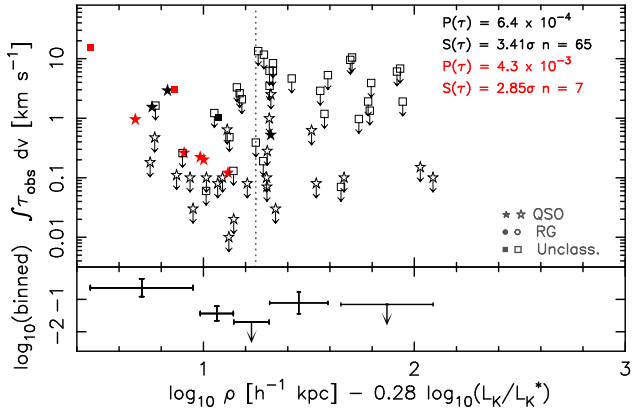}
\caption{As Fig. \ref{N-blue} but with the impact parameter normalised by the  K-band luminosity (cf. \citealt{clwb01}).}
\label{N-K}
\end{figure} 
Normalising the impact parameter, the significance of the anti-correlation increases (Fig. \ref{L-K}) and,
again comparing detection rates above and below the median (normalised) impact
parameter:  At $(\rho/h)\times(L_{\rm K}/L_{\rm K}^*)^{-0.28} <22.2$ kpc, there are 10 detections and 21 non-detections, that is
a 30.5 per-cent detection rate. Applying this to the $>22.2$ kpc bin, gives a binomial probability of
$3.05\times10^{-5}$ of obtaining one detection or fewer out of 33 sight-lines, which is significant at
$4.17\sigma$.

\begin{figure}
\centering \includegraphics[angle=-90,scale=0.48]{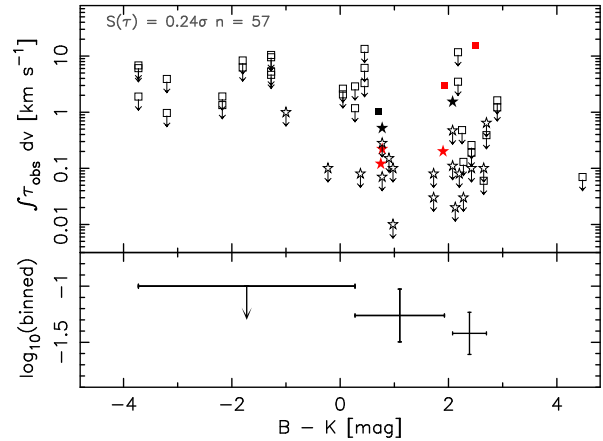}
\caption{The velocity integrated optical depth versus the blue--near-infrared colours of the sample.}
\label{colour}
\end{figure}

Since we have both the blue and near-infrared luminosities for many of the galaxies, we can verify that the observed
decrease in $\int\!\tau dv$ with impact parameter is not dominated by a colour bias, where the redder (and presumably,
dustier) galaxies are more hospitable to the presence of cool gas.\footnote{As is seen for both \HI\ \citep{cw10}
 and \MOLH\ \citep{cwc+11} at high redshift.}  In Fig. \ref{colour} we
plot the 21-cm absorption strength against the $B-K$ colour, where no correlation is apparent, leaving the decrease in gas density with impact parameter the major most likely cause.
 We also note that the known metal-line systems do exhibit some reddening ($B - K >0$), which is consistent with the
metallicity--dust (and molecular fraction) relation (see \citealt{cwmc03} and references therein).

\subsection{The effect of the background source size}
\label{ebss}

As mentioned in Sect. \ref{intro}, in addition to the impact parameter, it is believed that the extent of the background source can have an effect on the 21-cm detection
rate. Specifically, for a given absorption cross section\footnote{\HI\ 21-cm absorbing clouds in
external galaxies are believed to have 1.4~GHz cross-sections of $\sim100$ pc \citep{bra12,cag+13}.} different background source sizes will have
different observed optical depths, $\tau_{\rm obs}$, according to 
\begin{equation}
\tau \equiv-\ln\left(1-\frac{\tau_{\rm obs}}{f}\right) \approx\frac{\tau_{\rm obs}}{f} \text{  for  }\tau_{\rm obs}\lapp0.3  \Rightarrow \tau_{\rm obs}\approx f \tau,
\label{tau_eq}
\end{equation}
in the optically thin regime\footnote{The
  maximum of the sample is $\tau_{\rm obs}= 0.24$ \citep{zlp+15}.}, where the observed optical depth is defined as the ratio of the
line depth to the continuum ($\tau_{\rm obs}\equiv \Delta S/S$).  Given that the covering factor ranges over $0\leq f
\leq 1$ for zero to full coverage of the background flux, larger background sources could give systematically lower
observed optical depths.  That is, we may expect the more compact quasar sight-lines to give a higher detection rate than those
of radio galaxies.\footnote{\citet{cag+13} 
have shown that the observed anti-correlation between the atomic hydrogen column density and the projected linear size of the
background radio emission \citep{pcv03,gs06,gs06a,gss+06,omd06,css11} is most likely driven by 
the observed optical depth resulting from the coverage of the radio source.}

\begin{figure}
\centering \includegraphics[angle=-90,scale=0.6]{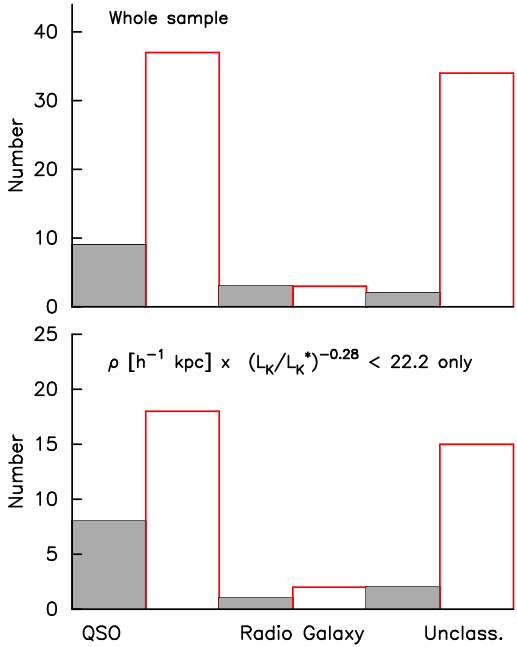}
\caption{Breakdown of the various sight-line types, from NED (where quasars are
  classified as QSOs), for the whole sample (top panel) and for impact parameters less than the median value (bottom). The
  filled histogram shows the 21-cm detections and the unfilled the non-detections.}
\label{sl}
\end{figure}

Addressing this, from the NED classifications, contrary to our expectations, we find a higher 
 ($50-100$ per-cent) detection rate towards radio galaxies compared to the quasars ($20-37$ per-cent, Fig.~\ref{sl}).
Performing a Pearson $\chi^2$--statistic test  and applying Yates's correction for continuity gives a significance of $p = 0.25$.
Thus, the null hypothesis that the detection rates are the equal for radio galaxies and quasars cannot be rejected.

It is generally accepted that the turnover frequency of a radio source is anti-correlated with its extent
(e.g. \citealt{ffs+90}), with flat spectral indices arising from radio emission projected along our line-of-sight, thus also
appearing more compact.  Therefore, in the absence of high resolution radio images, we can use any clustering of either
or both the turnover frequency/spectral index to gauge whether an unclassified background source is likely to be a
quasar or a radio galaxy. We therefore compiled the photometry of each background source and, using the methods
described in \citet{cwsb12}, determined the rest-frame turnover frequency, $\nu_{_{\rm TO}}$, and spectral index at a
rest frequency of 1.4~GHz, $\alpha$.

\begin{figure}
\centering \includegraphics[angle=-90,scale=0.5]{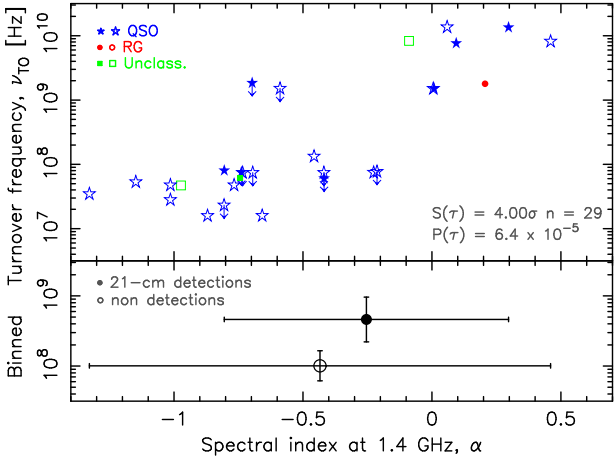}
\caption{The rest-frame turnover frequency versus the spectral index for the background sources for which these 
could be determined. If no turnover is apparent in the radio SED we assume that this occurs below the lowest
observed frequency (typically $\sim10$ MHz) and use this to assign an upper limit to $\nu_{_{\rm TO}}$.
Again, the filled symbols represent the detections and the unfilled the non-detections with shapes representing the
classification of the background source.
In the bottom panel, the binned values of the detections and non-detections are shown.}
\label{TO-SI}
\end{figure} 
From this (Fig.~\ref{TO-SI}), we see that both the radio galaxies and unclassified sources,
for which both $\nu_{_{\rm TO}}$ and $\alpha$ could be determined, are too few in number to exhibit any clustering, which is absent from the
sources classified as QSOs in any case. Thus, at least at present,
this method is of little use in predicting the nature of the background source. Based upon this limited sample, however,
we see that, while there is a large overlap in the spectral indices (although the detections are more ``flat'', $|\alpha| < 0.5$), the 
detections and non-detections have distinct mean turnover frequencies
($\nu_{_{\rm TO}} = 10^{8.66\pm0.32}$ and $10^{8.00\pm0.21}$ Hz, respectively). Since the turnover frequency is anti-correlated with the source size 
(e.g. \citealt{ode98,fan00,omd06}), this is consistent with a detection bias towards the more compact radio sources,  indicating that the covering factor is important.

\subsection{Spin temperature and impact parameter}
 
In the optically thin regime (Equ. \ref{tau_eq}), the neutral hydrogen column density, $N_{\text{\HI}}$ [\scm], is related to the absorption
strength via 
\begin{equation}
N_{\text{\HI}}  \approx 1.8\times10^{18}\,\frac{T_{\rm  spin}}{f}\int\!\tau_{\rm obs}\,dv,
\label{N_eq}
\end{equation}
where $T_{\rm spin}$ [K] is the spin temperature of the gas and $\int\!\tau_{\rm obs}\,dv$ [\kms] the velocity
integrated optical depth of the absorption. The comparison of $\int\!\tau_{\rm obs}\,dv$ with $N_{\text{\HI}}$, from
Lyman-\AL\ ($\lambda=1216$~\AA) observations, has shown an increase in $T_{\rm spin}/f$ with redshift, possibly
suggesting an evolution in the spin temperature (e.g. \citealt{kps+14}). However, the effect can adequately be accounted
for by a decrease in $f$ with redshift, due to the geometry effects of an expanding Universe \citep{cur12}.

\begin{figure}
\centering \includegraphics[angle=-90,scale=0.5]{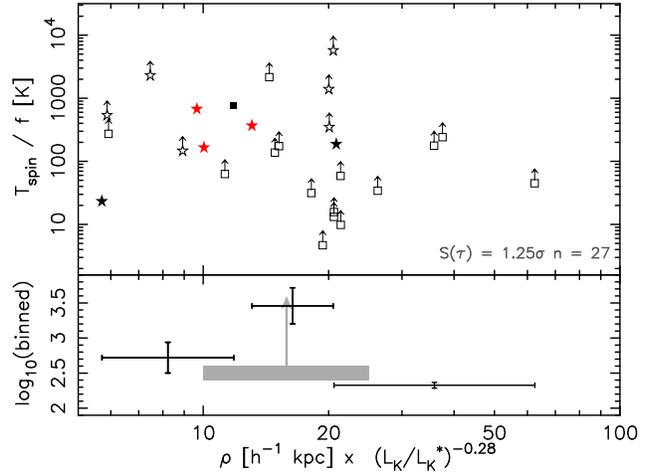}
\caption{The spin temperature--covering factor degeneracy versus the normalised impact parameter for all of the published searches
  where a measure of the column density is available. In the top panel, the symbols are as per Fig.~\ref{N-impact} and in the bottom panel
  the grey box shows the range in the outer Milky Way ($T_{\rm spin} = 250-400$ K, \citealt{dsg+09}). Since $f\leq1$, this places a lower
limit on $T_{\rm spin}/f$.}
\label{Toverf}
\end{figure}

For this near-by sample we have the opportunity to determine whether  $T_{\rm spin}/f$ generally varies with galactocentric radius, which
could possibly introduce a beam-filling effect, where the mean value of $T_{\rm spin}/f$ within the beam could also
contribute to a redshift dependence.  Given that the gas is optically thin, for  the low redshifts of the sample
(Fig.~\ref{distbn}), where 21-cm emission can be detected ($z\lapp0.2$, \citealt{cc15}), the column density can be
obtained directly from the brightness temperature of the line emission, $T_{\rm b}$ [K], via
\begin{equation}
N_{\text{\HI}}  \approx 1.8\times10^{18} \int\! T_{\rm b}\,dv,
\label{em_eq}
\end{equation}
which in conjunction with Equ. \ref{N_eq} yields $T_{\rm spin}/f$. Plotting this (Fig.~\ref{Toverf}), 
we see no evidence of a variation in $T_{\rm spin}/f$ with the normalised impact parameter over $(\rho/h)\times(L_{\rm K}/L_{\rm K}^*)^{-0.28} \lapp100$~kpc, which corresponds to  the inner $r\approx30$ kpc of galactocentric radius.

This is consistent the result of \citet{dsg+09}, who find remarkably  little variation in the spin temperature over 
radii of 8--25~kpc in the Milky Way, despite both the 21-cm emission and absorption strengths varying by
two orders of magnitudes.

\subsection{Disk inclination}

\begin{figure}
\centering \includegraphics[angle=-90,scale=0.52]{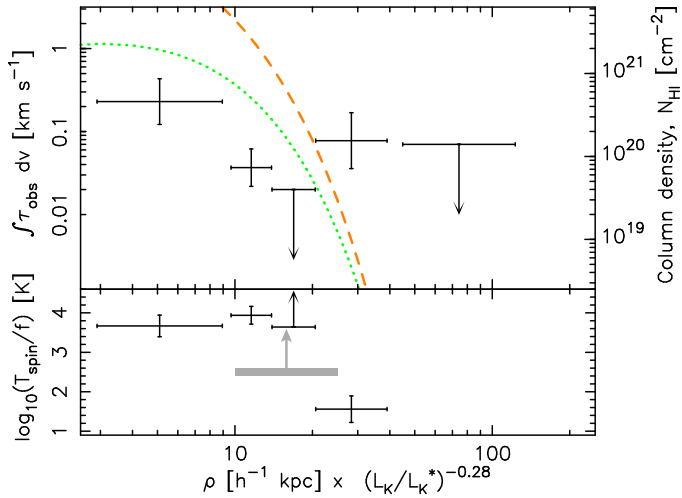}
\caption{Top: The binned 21-cm absorption strength (left ordinate axis, Fig.~\ref{N-impact}) versus the impact parameter
  overlaid with the Galactic \HI\ column density distribution (where $n_0 = 13.4$ \ccm, $R = 3$ kpc \& $f_{\rm FL} =20$,
  right ordinate axis, dotted curve -- viewed face-on, broken curve -- viewed edge-on). Bottom: The values of $T_{\rm
    spin}/f$ obtained from the face-on case. As per Fig. \ref{Toverf}, the grey box shows the range in the outer Milky
  Way.}
\label{model}
\end{figure}

By assuming the known \HI\ column density distribution of the Milky Way, we can estimate the approximate orientation of
the average \HI\ absorbing disk, based upon the $\int\tau dv$--$\rho$ relationship: The Galactic volume density of the
neutral gas exhibits an exponential decrease with galactocentric radius according to $n = n_0\,e^{-r/R}$, where $n_0 =
13.4$~\ccm\ and the scale-length $R = 3$~kpc \citep{kdkh07}.  The column density at each value of $\rho$ is obtained
from the volume density via $N_{\text{\HI}}\equiv\int\!ndl$, where $N_{\text{\HI}}= n\,h = n_0\,e^{-\rho/R}\,h$ for a
face-on ($i=90$\dg) disk\footnote{Where the height of the disk, $h$, is given by the flare factor, $f_{\rm FL} =
  \rho/h$, which describes the flaring of the \HI\ gas with galactocentric radius \citep{kdkh07}.} and $N_{\text{\HI}}=
n_0\int_{0}^{\rho}e^{-r/R}dr = n_0R\,e^{-\rho/R}$ for an edge-on ($i=0$\dg) disk (Fig. \ref{model}).

From this, the face-on galactic disk appears to better trace the binned integrated optical depth values, again with 
$T_{\rm spin}/f\sim10^3$~K 
remaining approximately consistent with galactocentric radius out to $(\rho/h)\times(L_{\rm K}/L_{\rm K}^*)^{-0.28} \sim20$ kpc,
beyond which this falls to $T_{\rm spin}/f=40$~K (Fig. \ref{Toverf}). 
Applying the
edge-on Galactic disk would give much larger values of $T_{\rm spin}/f$, due either to very high spin temperatures
and/or very low covering factors, the latter of which would be expected from a low inclination disk presenting a much
smaller absorbing cross-section.  We thus suggest that the targetted galaxies tend to have face-on inclinations, which
is consistent with the observations of \citet{rsa+15}, who, from the azimuthal profiles of the \HI\ emission, suggest
that the intervening galaxies could be highly inclined (or highly asymmetric).

\section{Conclusions}

One of the goals of current surveys for \HI\ 21-cm absorption at various impact parameters is to determine the expected
detection rate of cool, neutral gas in distant galaxies with the next generation of large radio telescopes. An important
aspect of this is the determination of the expected detection rate of 21-cm absorption at redshifts where the detection
of 21-cm emission will not be possible.  Here we demonstrate a clear anti-correlation between the
integrated optical depth and the impact parameter, where previously only weak correlations were apparent at best
(significance levels of $\leq0.81\sigma$, \citealt{gsb+10,bty+11,gsn+13,zlp+15}). 

Also, investigating the hypothesis that 21-cm absorption is more readily detected towards quasars in comparison to radio
galaxies, we find that the statistics are insufficient to draw any conclusions. However,  we do find that the mean detection occurs towards
a radio source with a higher turnover frequency than a non-detection. Since the turnover frequency is anti-correlated with
source size, this suggests that the coverage of the background emission is important and thus we would expect a higher detection
rate towards quasars in comparison to radio galaxies.

Lastly, using the neutral hydrogen column density obtained from the 21-cm emission, we find no statistical evidence for
a variation of the spin temperature (degenerate with the covering factor) over the inner $\approx30$ kpc of the disk,
similar to the consistency in spin temperature observed across the Milky Way \citep{dsg+09}.
By using $N_{\text \HI}$ obtained from the \HI\ distribution of the Milky Way, the observed 
$\int\tau dv$--$\rho$ correlation may suggest that the intervening galactic disks have high inclinations (close to face-on). This is consistent
with the \HI\ emission maps of \citet{rsa+15}, who suggest that galaxies which intercept background sources may be  highly inclined.

\section*{Acknowledgements}
We wish to thank the anonymous referee for their helpful and detailed comments and the {\em Radio Galaxy Morning Tea}
group at the University of Sydney for useful discussion.  This research has made use of the NASA/IPAC Extragalactic
Database (NED) which is operated by the Jet Propulsion Laboratory, California Institute of Technology, under contract
with the National Aeronautics and Space Administration and NASA's Astrophysics Data System Bibliographic Service. This
research has also made use of NASA's Astrophysics Data System Bibliographic Service and {\sc asurv} Rev 1.2
\citep{lif92a}, which implements the methods presented in \citet{ifn86}.


\label{lastpage}

\end{document}